# Perturbed Periodic Lattices: Sharp Crossover Between Effective-Mass-Like States and Wannier-Stark-Like Ladders

R. Merlin

*The Harrison M. Randall Laboratory of Physics, University of Michigan,  
Ann Arbor, Michigan 48109-1040, USA*

The concept of Wannier-Stark ladders, describing the equally spaced spectrum of a tightly- bound particle in a constant electric field, is generalized to account for arbitrary slowly-varying potentials. It is shown that an abrupt transition exists that separates Wannier-Stark-like from effective-mass-like behavior when the depth of the perturbation becomes equal to the width of the band of extended states. For potentials bounded from below, the spectrum bifurcates above the critical energy while the wavefunctions detach from the effective-mass region and split into two pieces.

The effective-mass approximation (EMA) allows one to replace the response of a particle in a periodic potential to a slowly-varying external perturbation with that of a particle possessing an effective mass, which is determined by the curvature of a band. In particular for semiconductors and conventional semimetals [1,2], the EMA significantly simplifies the calculation of a wide range of material properties, including exciton binding energies [3,4], cyclotron resonance [5], donor and acceptor levels [2] and carrier mobilities [6].

Fundamentally different that the EMA, Wannier-Stark (WS) ladders [7,8] are approximate solutions to the quantum problem of a particle subjected to both a periodic lattice potential and a constant electric field. The existence of a ladder of equally spaced energy levels is intimately related to the phenomenon of Bloch oscillations [9], which have been observed in, e. g., semiconductor superlattices [10] and accelerated optical lattices [11]. More recently, it has been shown that these oscillations play a central role in the generation of high harmonics in solids [12].

Even though the EMA and the occurrence of WS ladders [13] are well-established concepts, the relationship between these two models is far from clear and, to some extent, the reason why the electric field spectrum cannot be explained using the EMA is a bit puzzling. Here, we extend the notion of spectral ladders to encompass arbitrary potentials. Our analysis demonstrates that the simplest tight-binding model contains both ladders and EMA-like states and reveals the existence of a sharp boundary separating the two approximations.

For both the EMA and WS ladders to be applicable, it is essential that the perturbation be weak enough to disregard interband transitions [14]. We take this into account right from the beginning by considering the one-band tight-binding Hamiltonian

$$\hat{H} = -t\sum_{p=1}^{N-1}\left(|p\rangle\langle p+1| + |p+1\rangle\langle p|\right) + w\sum_{p=1}^{N} G(p)|p\rangle\langle p| \quad , \tag{1}$$

which describes a single particle in a lattice of $N$ sites. Here, $|p\rangle$ denotes the state with the particle at site $p$, with $\langle i|j\rangle = \delta_{ij}$, $t > 0$ is the nearest-neighbor hopping, $w > 0$ is a constant and $G(p)$ is a slowly-varying function of the site index. In the following, we assume that the external potential has a single minimum at $p = p_0$ and, without loss of generality, that $G(p_0) = -1$. The eigenfunctions of $\hat{H}$ are of the form

$$\Psi = \sum_p \Phi_p |p\rangle \qquad (2)$$

where the coefficients $\Phi_p$ satisfy the secular equation

$$[-E + wG(p)]\Phi_p = t(\Phi_{p-1} + \Phi_{p+1}) \ . \qquad (3)$$

For $w = 0$, the solutions for rigid wall boundary conditions are the extended states $\Phi_p \propto \sin(mp\pi/N)$ with $E_m = -2t\cos(m\pi/N)$; $m = 1,..,N$.

The EMA assumes that the coefficients do not vary much across neighboring cells. This allows for the treatment of the site index as a continuous variable, that is, $p$ becomes $\xi = x/a$ where $x$ is the space coordinate and $a$ is the lattice parameter. Adding $-2t\Phi_p$ to both sides of Eq. (3), the right-hand side can be written as $\approx t\, d^2\Phi/d\xi^2$. Thus, the expression becomes

$$-t\frac{d^2\Phi}{d\xi^2} + [wG(\xi) - 2t]\Phi = E\Phi \ , \qquad (4)$$

which is the Schrödinger equation of a particle of effective mass $1/2t$ in the potential $wG(\xi) - 2t$. Highlighting the importance of the assumption that $G$ is slowly varying, it should be noted that this does not necessarily ensure the same for $\Phi_p$. Hence, the validity of the EMA relies on the expectation that Eq. (4) has solutions that exhibit slow variation. Let $\alpha^{-1} \gg 1$ be the characteristic scale of $G$. Near the minimum at, say, $\xi = \xi_0$, $G \approx -1 + \alpha^2(\xi - \xi_0)^2$. Using results for the harmonic oscillator, and ignoring factors of order one, the scale for $\Phi$ is thus $\ell \sim (t/\alpha^2 w)^{0.25}$, which has to be large for the EMA to apply. As discussed later, this is a necessary but insufficient condition for its validity.

Returning to Eq. (3), we now consider the limit when the hopping term can be treated as a perturbation so that, to lowest order, we have a ladder of eigenenergies $E_j^{(0)} = w\,G(j)$. We will denote this limit as the potential-ladder approximation (PLA). Expanding $G$ around $p = j$, we get

$$w\left[G'(j)(p-j)+...\right]\Phi_p = t(\Phi_{p-1} + \Phi_{p+1}) . \tag{5}$$

If the nonlinear terms in the expansion can be ignored, that is, if $|G'| \gg \sqrt{|(t/w)G''|}$, this equation has the exact solution

$$\Phi_p^{(j)} = J_{p-j}(2\eta) \tag{6}$$

where $\eta = t/wG'(j)$. With the origin at $p = j$, $\left|\Phi_p^{(j)}\right|$ is an even function that decays exponentially outside the interval $|p-j| \leq \pi\eta$. It is obvious that the approximation becomes exact for the problem of a constant field where the eigenvalue spectrum is the equally-spaced WS ladder [7] and, moreover, that the validity of the PLA does not depend on the length scale of $G$. Because the Bessel function varies rapidly around $p = j$, it is also clear that Eq. (6) does not meet the EMA requirements.

Figure 1 shows a small set of the bound eigenstates obtained from numerically solving Eq. (3) for the Morse potential

$$G(p) = e^{-2\alpha(p-p_0)} - 2e^{-\alpha(p-p_0)} \tag{7}$$

with minimum at $p = p_0$. The data in the top and bottom insets demonstrate that the eigenfunctions at energies for which $|G'|$ is large and near the minimum of $G$ are accurately described, respectively, by the expressions corresponding to the PLA, Eq. (6), and the EMA, Eq. (4).

In order to understand the general properties of the localized eigenstates, consider the transfer matrix relating a pair of neighboring sites to one of its adjacent pairs. From Eq. (3), and using the fact that $G$ is slowly varying, we get

$$\begin{pmatrix} \Phi(p-2) \\ \Phi(p-1) \end{pmatrix} \approx \begin{bmatrix} a^2 - 1 & -a \\ a & -1 \end{bmatrix} \begin{pmatrix} \Phi(p) \\ \Phi(p+1) \end{pmatrix} \quad , \qquad (8)$$

where $a = [-E + wG(p)]/t$. The eigenvalues of this matrix are $(a^2 - 2 \pm a\sqrt{a^2 - 4})/2$, and their character determines the behavior of the wavefunctions. A real eigenvalue at a particular site, that is, a site for which $|a| \geq 2$, indicates that the state decays or grows exponentially around the site. If, instead, $|a| < 2$, the eigenvalue is a complex number of unit modulus and, therefore, one expects the wavefunction to vary weakly from one pair to its neighbors. These considerations indicate that the solutions to the secular equation, Eq. (3), satisfy the inequality

$$|-E + wG(p)| < 2t \quad . \qquad (9)$$

For bound states, this inequality defines the region within which a given state is confined. Leaving aside the question of the suitability of the EMA and PLA, the above exact expression must be clearly obeyed by all the eigenstates, including those for which the two approximations apply. Not surprisingly, the localization behavior of the bound eigenstates depicted in Fig. 1 is in excellent agreement with the inequality.

We now show that the assumption that the external potential has a minimum leads to the bifurcation of the wavefunction beyond a certain threshold energy $E_C$ so that for $E \leq E_C$ and $E > E_C$, the eigenstates are, respectively, EMA-like and PA-like. By PA-like, we mean states like the blue ones in Fig. 1 for which the probability to find the particle in a neighborhood of the minimum is exponentially small while by EMA-like, we refer to wavefunctions like the red one with a high probability to find the particle in the vicinity of $p = p_0$. Using that $G(p_0) = -1$ and Eq. (9), it follows that EMA-like states exist for $-(w+2t) < E < -w+2t$ and, therefore, that $E_C = -w + 2t$. As shown in Fig. 1, for eigenenergies above the threshold, the wavefunction detaches from the minimum as it splits into two pieces. Although not readily apparent in the linear scale of the figure, it is worth mentioning that a PA-like eigenstate with energy $E_j$ reemerges on the opposite side of the potential minimum, near the site where $G \approx E_j / w$. It should be emphasized that these findings are not restricted to the Morse potential exclusively but extend to arbitrary perturbations with $G(p) > -1$. In the particular case where the external potential is even with respect to $p = p_0$, the PA-like eigenstates are nearly-degenerate combinations of the form $\chi(p_0 - p) \pm \chi(p - p_0)$ that are localized on sites at both sides of the minimum. If, in addition, $\lim_{p \to \infty} G(p) = 0$, the necessary condition for the existence of PA-like eigenstates is $w > 4t$ because the onset of the continuum is at $E = -2t$. It is also important to note that, even in the case where $\ell \sim (t / \alpha^2 w)^{0.25} \gg 1$, the EMA is no longer valid for $w > 4t$.

Figure 2 reveals that the eigenenergy spectrum displays a bifurcating pattern similar to that of the wavefunctions. Below $E_C$, the data show a single branch that, at low energies, converges towards the expression for the EMA free-particle eigenvalues [15]

$$E_n = -w\left[1 - \sqrt{\frac{\alpha^2 t}{w}}(n+\frac{1}{2})\right]^2 \quad , \tag{10}$$

represented by the red curve. Above the critical energy, there are two branches (diamonds and circles) which are very well described by the PLA, that is, $E_j = w\, G(j)$ (blue curve). The transition between the two regimes manifests itself as a dip in the data of the inset, which shows the difference between two consecutive eigenvalues. Also note the dip at the boundary with the continuum at $E = -2t$.

In conclusion, we have shown that bound eigenstates of the tight-binding model divide into two qualitatively distinct sets that resemble EMA states at energies close to the minimum of the external potential and PLA states above a critical energy that sharply separates the two groups. The existence of PLA-like states requires that (*i*) the depth of the potential be larger than the width of the band [16], and that (*ii*) effects of interband transitions can be ignored (as in the case of a constant electric field, interband coupling is expected to convert the PLA-like states into resonances with lifetimes determined by the strength of interband tunneling [17]). While the observation of such states and, more importantly, the EMA-to-PLA transition are challenging in bulk materials due to their wide bandwidths, the effective narrow widths that can be attained in artificial systems like semiconductor superlattices [13], two-dimensional heterostructures [18], optical lattices [11], and twisted graphene [19] provide promising avenues for their discovery.

FIGURE CAPTIONS

FIGURE 1 – Results of calculations for $N = 10^3$, $\alpha = 0.01$, $w = 16$ and $t = 1$. The black curve is the Morse potential, Eq. (7). The red and blue curves are linear plots of the magnitude of the eigenfunctions of relative energies (left to right) $E/t$ = 1.15, -10.7, -17.12, - 9.517 and - 2.085. Red and blue denote, respectively, EMA- and PLA-like states. Top inset: Comparison between the eigenfunction obtained numerically from the secular equation (blue circles) and the PLA expression (black curve), Eq. (6), for the eigenstate with $E/t$ = -10.105. Bottom inset: Comparison involving the eigenstate of relative energy $E/t$ = -17.193 (red circles) and the EMA (black curve), Eq. (4).

FIGURE 2 – Eigenenergies near the EMA-PLA crossing calculated for $N = 10^3$, $\alpha = 0.01$, $w = 16$ and $t = 1$. The two branches of the spectrum are represented by diamonds and circles. A different level ordering was used to allow for a comparison between the two solutions above $E_C$ and the Morse potential (blue curve); for this reason, only one of them connects smoothly with the single solution below the critical energy. The red curve is the free particle expression for the eigenenergies, Eq. (10). Inset: $\Delta E$ is the energy difference between two consecutive eigenenergies for the branch labeled with circles. Vertical lines depict the EMA-PLA boundary at $E = -w + 2t$ and the PLA-continuum boundary at $E = -2t$. The continuum is in the region $-2t < E < 2t$, and the overall spectrum extends from $-(w + 2t)$ to $2t$.

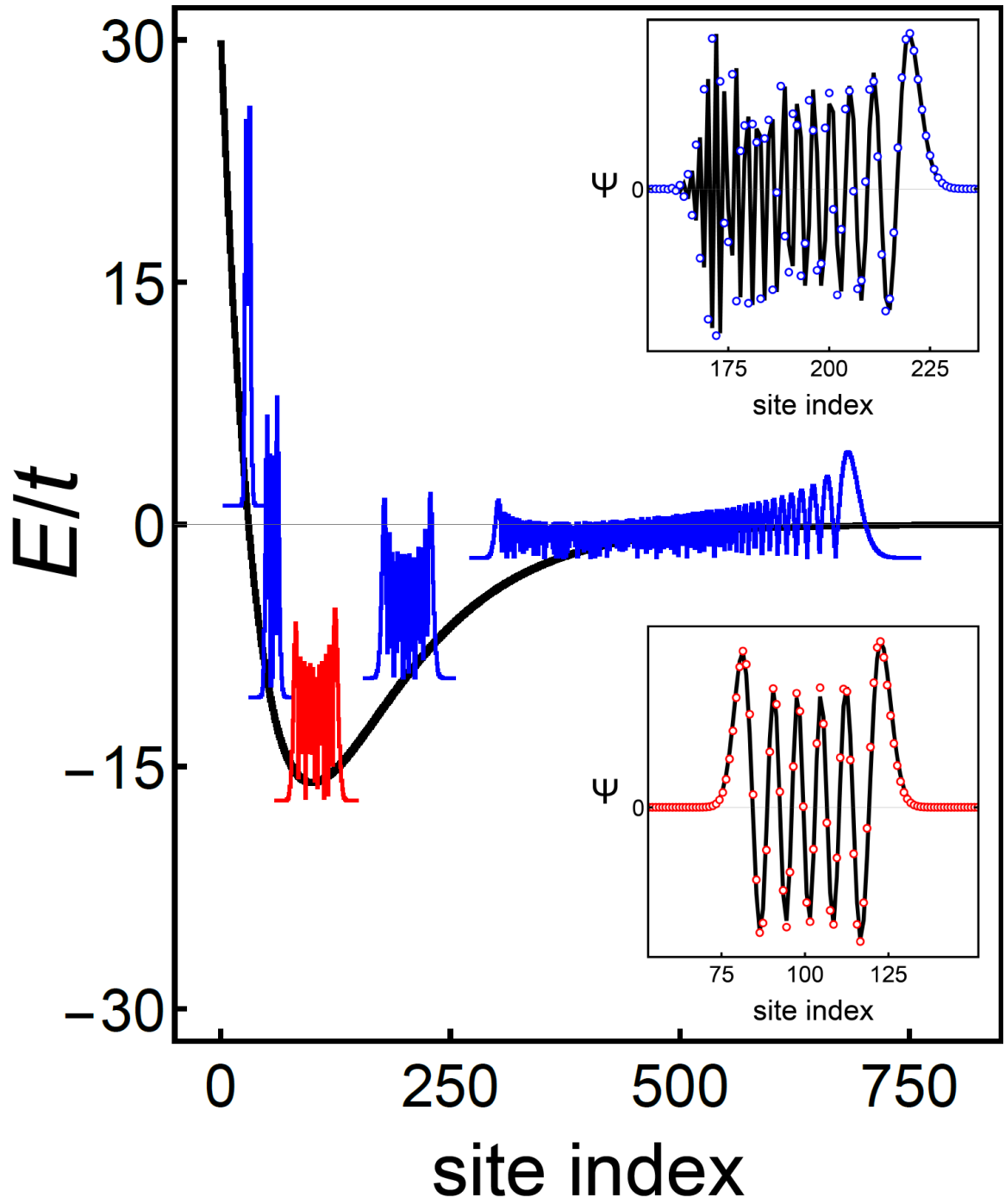

FIGURE 1

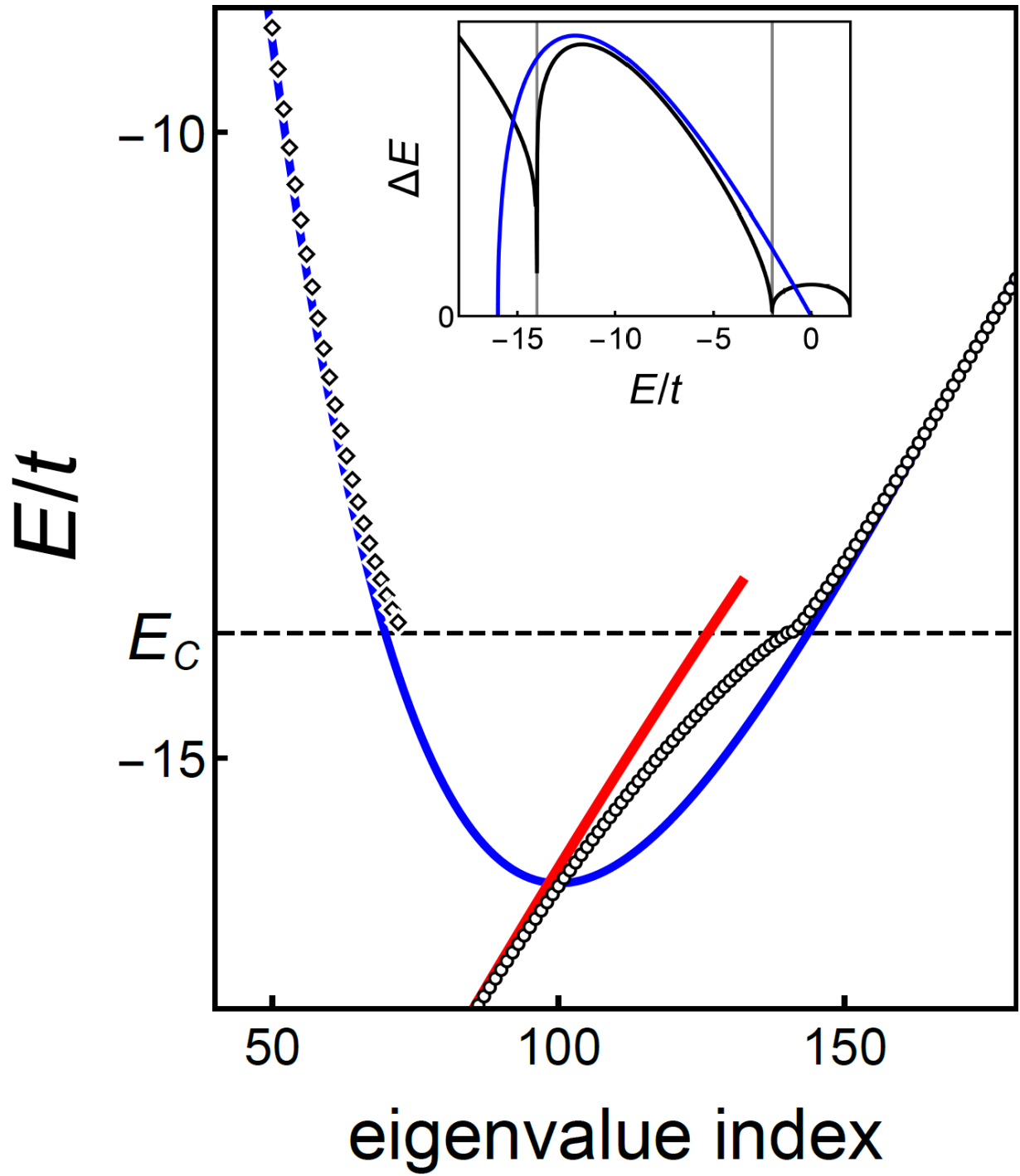

FIGURE 2